\documentclass[journal=jacsat,manuscript=article]{achemso}
\pdfoutput=1
\usepackage[version=3]{mhchem} % Formula subscripts using \ce{}
\usepackage[]{mathrsfs}
\usepackage[]{listings}
\usepackage[latin1]{inputenc}
\usepackage{amssymb}
\usepackage{epsfig}
\usepackage{xspace}
\usepackage{pstricks}
\usepackage{psfrag}
\usepackage{textgreek}

\newcommand{\Fig}[1]     {Fig.~\ref{#1}}

\newcounter{myitem}

\newcommand{\resetmycnt}{\setcounter{myitem}{1}}

\resetmycnt

\newcommand{\figdir}  {./figures/paper}

\newcommand*{\citen}[1]{%
  \begingroup
    \romannumeral-`\x % remove space at the beginning of \setcitestyle
    \setcitestyle{numbers}%
    \cite{#1}%
  \endgroup   
}
\author{\vspace*{-0.1cm}\it Yury Yu. Illarionov}
\affiliation{Institute for Microelectronics (TU Wien), Gusshausstrasse 27--29, 1040 Vienna, Austria}
\email{illarionov@iue.tuwien.ac.at}
\alsoaffiliation{Ioffe Physical-Technical Institute, Polytechnicheskaya 26, 194021 St-Petersburg, Russia}
\author{\vspace*{-0.1cm}\it Alexander G. Banshchikov}
\affiliation{Ioffe Physical-Technical Institute, Polytechnicheskaya 26, 194021 St-Petersburg, Russia}
\author{\vspace*{-0.1cm}\it Dmitry K. Polyushkin}
\affiliation{Institute for Photonics (TU Wien), Gusshausstrasse 27--29, 1040 Vienna, Austria}
\author{\vspace*{-0.1cm}\it Stefan Wachter}
\affiliation{Institute for Photonics (TU Wien), Gusshausstrasse 27--29, 1040 Vienna, Austria}
\author{\vspace*{-0.1cm}\it Theresia Knobloch}
\affiliation{Institute for Microelectronics (TU Wien), Gusshausstrasse 27--29, 1040 Vienna, Austria}
\author{\vspace*{-0.1cm}\it Mischa Thesberg}
\affiliation{Institute for Microelectronics (TU Wien), Gusshausstrasse 27--29, 1040 Vienna, Austria}
\author{\vspace*{-0.1cm}\it Michael St\"{o}ger-Pollach}
\affiliation{University Service Center for Transmission Electron Microscopy (TU Wien), Wiedner Hauptstrasse 8-10/052, 1040 Vienna, Austria}
\author{\vspace*{-0.1cm}\it Andreas Steiger-Thirsfeld}
\affiliation{University Service Center for Transmission Electron Microscopy (TU Wien), Wiedner Hauptstrasse 8-10/052, 1040 Vienna, Austria}
\author{\vspace*{-0.1cm}\it Mikhail I. Vexler}
\affiliation{Ioffe Physical-Technical Institute, Polytechnicheskaya 26, 194021 St-Petersburg, Russia}
\author{\vspace*{-0.1cm}\it Michael Waltl}
\affiliation{Institute for Microelectronics (TU Wien), Gusshausstrasse 27--29, 1040 Vienna, Austria}
\author{\vspace*{-0.1cm}\it Nikolai S. Sokolov}
\affiliation{Ioffe Physical-Technical Institute, Polytechnicheskaya 26, 194021 St-Petersburg, Russia}
\author{\vspace*{-0.1cm}\it Thomas Mueller}
\affiliation{Institute for Photonics (TU Wien), Gusshausstrasse 27--29, 1040 Vienna, Austria}
\author{\vspace*{-0.1cm}\it Tibor Grasser}
\affiliation{Institute for Microelectronics (TU Wien), Gusshausstrasse 27--29, 1040 Vienna, Austria}
\email{grasser@iue.tuwien.ac.at}

\title[An \textsf{achemso} demo]
  {Meeting the Scaling Challenge for Post-Silicon Nanoelectronics using CaF$_\textbf{2}$ Insulators}

\begin{document}



\begin{abstract}
Two-dimensional (2D) semiconductors have been suggested both for ultimately-scaled field-effect transistors (FETs) and More-than-Moore nanoelectronics. However, these targets can not be reached without accompanying gate insulators which are scalable into the nanometer regime. Despite the considerable progress in the search for channel materials with high mobilities and decent bandgaps, finding high-quality insulators compatible with 2D technologies has remained a challenge. Typically used oxides (e.g. SiO$_2$, Al$_2$O$_3$ and HfO$_2$) are amorphous when scaled, while  two-dimensional hBN exhibits excessive gate leakages. To overcome this bottleneck, we extend the natural stacking properties of 2D heterostructures to epitaxial fluorite (CaF$_2$), which forms a quasi van der Waals interface with 2D semiconductors. We report scalable single-layer MoS$_2$ FETs with a crystalline CaF$_2$ insulator of about 2$\,$nm thickness, which corresponds to an equivalent oxide thickness of less than 1$\,$nm.  While meeting the stringent requirements of low leakage currents, our devices exhibit highly competitive performance and record-small hysteresis. As such, our results present a breakthrough for very large scale integration towards commercially competitive nano-electronic devices.   

\end{abstract}
%\textbf{Keywords:} Black phosphorus, MoS$_2$, graphene, transistor, 2D electronics, reliability improvement


%Fabrication of different electronic devices with atomically thin two-dimensional (2D) layers has achieved a considerable progress in the past few years. In particular, these emerging technologies have been 
%suggested for applications in flexible electronics~\cite{LEE13,AKINWANDE14}, optoelectronic devices~\cite{AVOURIS14,BABLICH16} and sensors~\cite{SHAVANOVA16}. Also,
Various two-dimensional (2D) semiconductors are now considered as channel materials in next-generation field-effect transistors (FETs), which are potentially suitable to extend the life of Moore's law by enabling scaled
channel geometries below 5$\,$nm. For instance, several groups have reported FETs with graphene~\cite{GUERRIERO17}, silicene~\cite{TAO15}, black 
phosphorus~\cite{LI14B,CHEN18} and transition metal dichalcogenides, such as  MoS$_2$~\cite{KANG14B,CHUANG14,ILLARIONOV17C,SMITHE17A,BOLSHAKOV17}, 
MoSe$_2$~\cite{LIAO17}, MoTe$_2$~\cite{CHO18}, WS$_2$~\cite{YANG14Y} and WSe$_2$~\cite{LIU13A,PRAKASH17}. Excellent transistor characteristics have already been obtained for MoS$_2$ FETs, 
such as on/off current ratios~\cite{ILLARIONOV17C} up to 10$^{10}$ and sub-threshold swing values down to 69$\,$mV/dec~\cite{BOLSHAKOV17}. Furthermore, some attempts on circuit integration of MoS$_2$ FETs have already been undertaken~\cite{WANG12P,DAS16}. 

However, one of the major research problems of 2D technologies is their miniaturization without considerable loss in already achieved device performance thresholds. While fabrication of competitive 2D FETs with scaled channel dimensions is already possible~\cite{XIE17}, scaling and precise control of the insulator thickness and quality remain a challenge. Typically, oxides known from Si technologies (e.g. SiO$_2$, Al$_2$O$_3$ and HfO$_2$) are used. These materials are amorphous when grown in thin layers, which makes the fabrication of high-quality interfaces with the channel difficult. In order to address this problem, different insulators have to be identified for next-generation 2D technologies. In particular, hexagonal boron nitride (hBN) has been intensively discussed in the literature~\cite{HUI16}. However, hBN has a small bandgap of about 6$\,$eV~\cite{CASSABOIS16}, a small dielectric constant of 5.06~\cite{GEICK66}, and unfortunate band offsets to most 2D materials. As scaled technologies require equivalent oxide thicknesses (EOT) below 1$\,$nm (corresponding to a physical thickness of below 1.3$\,$nm in hBN), hBN will result in excessive thermionic and tunneling leakage currents (see Supplementary Section 1). To overcome this bottleneck, we suggest here to use epitaxially grown (and thus crystalline) calcium fluoride (fluorite, CaF$_2$) which has outstanding insulating properties even for a physical thickness of just about 2$\,$nm (EOT of about 0.9$\,$nm).  
%, and breakdown fields barely exceeding 4$\,$MV/cm~\cite{JANG16}. For example, the MoS$_2$ FETs with the thinnest insulator reported so far~\cite{BOLSHAKOV17}, have a top gate HfO$_2$ insulator of 4$\,$nm 
%and a back gate Al$_2$O$_3$ as thick as 15$\,$nm, which is considerably thicker than the insulators with sub-1$\,$nm equivalent oxide thicknesses (EOT) routinely available in standard silicon technology.

Epitaxially grown CaF$_2$ is a high-k crystalline material with a very favourable combination of dielectric properties~\cite{HAYES74}, such as a high dielectric constant ($\mathrm{\varepsilon}\,=\,$8.43), wide bandgap ($E_{\mathrm{g}}$\,=\,12.1$\,$eV) and high effective carrier mass ($m^*$\,=\,1.0$m_0$). Also, its nearly perfectly matching lattice constant (0.546$\,$nm) with Si (0.543$\,$nm) and similarities between the fluorite and Si lattice 
structures allow growth of CaF$_2$ layers on Si and Ge substrates using molecular-beam epitaxy (MBE)~\cite{SUGIYAMA96,ILLARIONOV14D} with very high quality. Furthermore, the CaF$_2$(111) surface is  
terminated by F atoms, which makes it chemically inert and free of dangling bonds with H passivation~\cite{FOSTER09} known to result in numerous reliability challenges. Owing to this, heteroepitaxy of 2D materials on the 3D CaF$_2$ surface is possible even for considerable lattice mismatch~\cite{KOMA90}, while leading to high-quality quasi van der Waals interface. In particular, recent studies have reported epitaxy of MoSe$_2$~\cite{VISHWANATH15} and MoTe$_2$~\cite{VISHWANATH18} on CaF$_2$(111) bulk substrates. This further underlines that CaF$_2$ is fully compatible with 2D semiconductors and thus can be considered an extremely promising insulating material for very large scale integration of 2D devices. 

Note that CaF$_2$ is only one material out of a wide class of epitaxial fluorides~\cite{SUGIYAMA96}. Some of these materials (e.g. anti-ferromagnetic NiF$_2$~\cite{BANSCHIKOV15} and MnF$_2$~\cite{KAVEEV05}, diamagnetic ZrF$_2$~\cite{KAVEEV05}, ferroelectric BaMgF$_4$~\cite{RAVEZ97}) have additional fascinating properties which may revolutionize future electronic device technologies. Nevertheless, the real potential of fluorides in modern electronic devices is still not fully exploited. For instance, previously, CaF$_2$ films have been mostly used as barrier layers in resonant tunneling diodes~\cite{WATANABE00} and superlattices~\cite{SUTURIN00}, together with CdF$_2$ films. As for the use of CaF$_2$ as a gate insulator in FETs, only a few working transistors with 640$\,$nm thick CaF$_2$ and poor reproducibility were reported~\cite{SMITH84}. The considerable progress in MBE grown tunnel-thin CaF$_2$ layers achieved in the last decade has revived the idea of fluoride insulators in FETs~\cite{TYAGINOV14}.
%Similarly to synthesis of 2D materials, the technology %of CaF$_2$ epitaxy has passed a long way towards considerable improvement.
%At the current state of the art, a homogeneous high-quality CaF$_2$ layer can be grown using MBE on silicon substrates with (111) orientation. In addition to the excellent dielectric properties of CaF$_2$ and well defined CaF$_2$(111) surface, this material combination also contributes to a decrease of the tunnel leakage current since tunneling through crystalline CaF$_2$ on Si(111) conserves transverse momentum, which leads to a smaller tunneling 
%probability and thus lower leakage currents~\cite{VEXLER10A,ILLARIONOV11C}. This is a considerable advantage over amorphous high-k oxides~\cite{ILLARIONOV10} which are typically grown on Si(100), not to mention 
%hBN which is sometimes even considered a wide bandgap semiconductor~\cite{CASSABOIS16} rather than an insulator. Subsequent adjustment of Si surface preparation and MBE growth parameters allowed us to achieve ultra-thin (3$-$7$\,$monolayers (ML), 1$\,$ML$\,$=$\,$0.315$\,$nm) CaF$_2$ films on Si(111) with minor thickness fluctuations and an overall quality which is high enough for device applications~\cite{ILLARIONOV14D}. In addition, 
%some successful attempts to fabricate high-quality CaF$_2$ layers on Ge(111) substrates\cite{ILLARIONOV11C} have been undertaken.
%However, the practical realization of CaF$_2$ dielectrics
%has become possible only recentlynow, owing to an intensive development of 2D technologies which requires beyond-CMOS materials.

In this study we combine the epitaxial growth of CaF$_2$ with chemical vapour deposition (CVD) of MoS$_2$ in a single device. In order to study the variability and reproducibility of these devices, we 
fabricated hundreds of electrically stable single-layer MoS$_2$ FETs with about 2$\,$nm thick CaF$_2$ gate insulators. Ultra-thin CaF$_2$ layers were deposited onto an atomically clean Si(111) surface using our 
well-adjusted MBE growth technique~\cite{ILLARIONOV14D} at 250$^{\mathrm{o}}$C (for more details see the Methods section). The growth process and crystalline quality of CaF$_2$ were 
controlled in real time using reflection high-energy electron diffraction (RHEED)~\cite{SOKOLOV92}, the corresponding images are provided in Supplementary Section 2. The thickness of CaF$_2$ measured by a quartz oscillator 
is 6$\,$--$\,$7$\,$ monolayers (ML, 1$\,$ML$\,$=$\,$0.315$\,$nm), which is close to the values measured using transmission electron microscope (TEM). The first step of transistor fabrication consisted in formation of SiO$_2$(5$\,$--$\,$10$\,$nm)/Ti/Au source/drain contact pads using sputtering. Then, a single-layer MoS$_2$ film grown by CVD~\cite{DUMCENCO15} at 750$^{\mathrm{o}}$C was transferred onto the substrate according to the method of~[\citen{GURARSLAN14}] (for more details see the Methods section) and MoS$_2$ channels with $L$ and $W$ varied between 400 and 800$\,$nm were shaped. Finally, additional e-beam evaporated Ti/Au layers were deposited to contact the channels. More details about the structure of our devices can be found in Supplementary Section 3. 

\begin{figure}[!h]
\vspace{0mm}
\begin{minipage} {\textwidth} %{\sminipagewidth}
\hspace{1cm}
  \includegraphics[width=14.5cm]{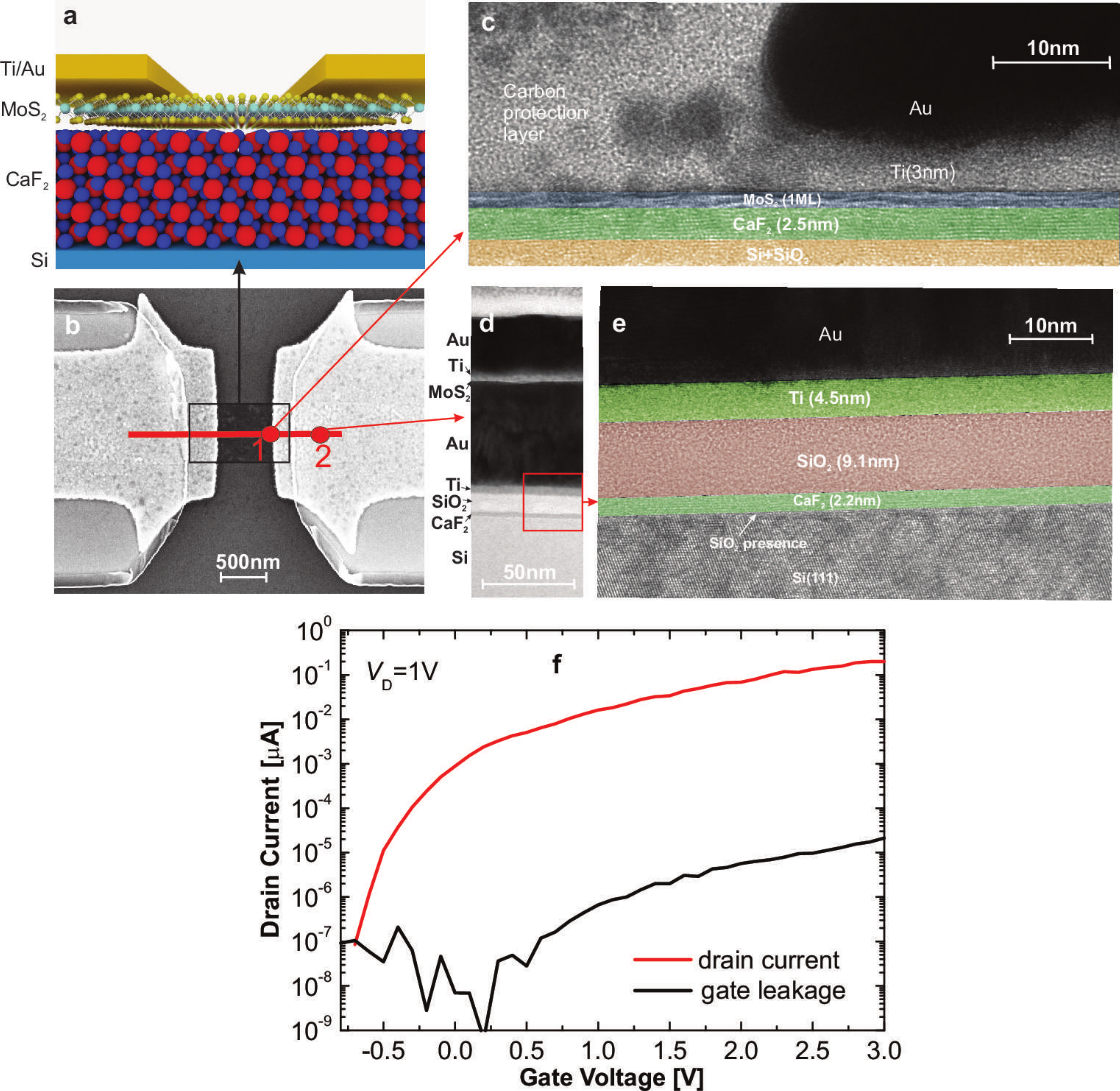} %???
\caption{\label{Fig.1} (a) Atomistic structure of the quasi van der Waals interface between F-terminated CaF$_2$(111) and MoS$_2$ in the channel area of our devices. (b) SEM image of our MoS$_2$ FET. The black box indicates the channel area with the source/drain electrodes, where MoS$_2$ is on top of CaF$_2$, as shown in (a). The red line indicates the approximate location of the cut for the TEM sample, with two locations where the images were collected. (c) TEM image obtained in the channel area (location 1). (d) Low resolution TEM image obtained ouside the channel area (location 2). The structure is the same as for the contact pads, with an SiO$_2$ isolation layer deposited on top of CaF$_2$ and MoS$_2$ sandwiched between two metal layers. (e) High resolution TEM image obtained in location 2 where the CaF$_2$ layers are visible. (f) Gate leakage through the CaF$_2$ layer is negligible compared to the drain current, which underlines the high quality of our MoS$_2$ FETs.}
\end{minipage}
\end{figure}

In \Fig{Fig.1}a we show that the atomic structure of CaF$_2$(111) is rather similar to that of 2D materials, with F-Ca-F monolayers having a thickness of 0.315$\,$nm. This makes fluorite a natural candidate for the integration into 2D process flows. An essential ingredient of our devices is the virtually defect-free CaF$_2$(111)/MoS$_2$ interface, which is formed by the F-terminated fluorite substrate, a quasi van der Waals gap and an atomically flat MoS$_2$ layer (\Fig{Fig.1}a). This interface is present in the channel area and under the source/drain electrodes, as marked in the scanning electron microscope (SEM) image in \Fig{Fig.1}b. In order to verify the layer structure of our device, we cut a 70$\,$nm thick specimen using focused ion beam (FIB) along the line marked in \Fig{Fig.1}b and performed TEM measurements. In \Fig{Fig.1}c we show a TEM image obtained for the channel area. We can clearly see the interface between single-layer MoS$_2$ and layered crystalline CaF$_2$ of about 8$\,$ML, which corresponds to a physical thickness of about 2.5$\,$nm. For different substrates, the number of CaF$_2$ monolayers varies between 6 and 8 (thickness between 1.9 and 2.5$\,$nm). By recording electron energy loss spectra (EELS, see Supplementary Section 5) at the interface between CaF$_2$ and the Si substrate we observe some SiO$_2$ (less than 1$\,$nm thick), which is formed by oxidation resulting from prolonged exposure to air of our CaF$_2$/Si substrates before device fabrication. Taking into account thickness fluctuations~\cite{ILLARIONOV14D} of CaF$_2$ and the presence of a thin thermal oxide layer, we model the tunnel leakages measured for numerous devices and find that the effective gate insulator thickness is about 2$\,$nm (see Supplementary Section 4). The layered structure of our CaF$_2$ films is clearly visible in the TEM image obtained using low dose imaging. Although electron irradiation is known to be destructive for CaF$_2$ samples~\cite{JIANG12}, these investigations indicate the extremely high stability of our thin CaF$_2$ layers, where the desorption of F by the electron beam and subsequent formation of CaO are not as favourable as in CaF$_2$ bulk crystals~\cite{JIANG12}. Nevertheless, we found that TEM measurements can destroy the Si substrate a few nanometers below the Si/CaF$_2$ interface, which, however, only occurs within the channel area (see Supplementary Section 5). Interestingly, outside the channel area our sample is unaffected by TEM irradiation. There we can clearly see MoS$_2$ sandwiched between two metal layers and an SiO$_2$ isolation layer on top of the CaF$_2$/Si substrate (\Fig{Fig.1}d,e). As a final verification of the properties of our 2$\,$nm thick CaF$_2$ insulator, in \Fig{Fig.1}f we show that the measured gate leakage is negligible compared to the drain current of our MoS$_2$ FET. 

%In particular, it is known that O impurities can be present in CaF$_2$ layers or substitute S vacancies in MoS$_2$, while their number can considerably increase after unavoidable ambient exposure of thin ($\sim$68$\,$nm) TEM specimen. Under electron irradiation, these O impurities migrate towards the edge of TEM specimen and cause oxidation~\cite{JIANG12}, which appears as an artifact in TEM images (\Fig{Fig.1}b). Remarkably, for our best devices this TEM-caused oxidation mostly appears in the interfacial layer of Si rather than in CaF$_2$ itself. At the same time, in the contact areas covered by gold oxidation is weakly pronounced or completely absent (\Fig{Fig.1}c-d), likely because of less intensive O penetration. While the physical thickness of CaF$_2$ varies between 1.9 and 2.5$\,$nm, the effective thickness of gate insulator should be identified taking into account possible thickness fluctuations of CaF$_2$ films and oxidation of thinner places.  

\begin{figure}[!h]
\vspace{0mm}
\begin{minipage} {\textwidth} %{\sminipagewidth}
\hspace{0cm}
  \includegraphics[width=16.0cm]{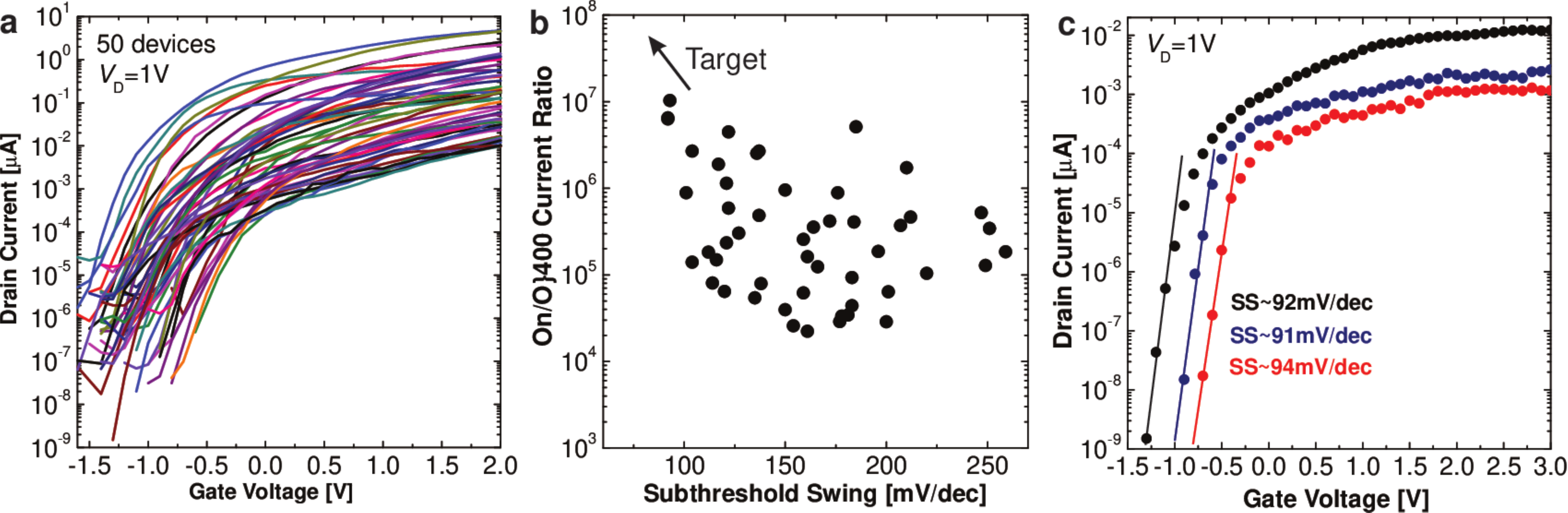} %???
\caption{\label{Fig.2} (a) $I_{\mathrm{D}}-V_{\mathrm{G}}$ characteristics of 50 MoS$_2$/CaF$_2$ FETs from two different Si/CaF$_2$ substrates. (b) Distributions of measured on/off current ratios and subthreshold 
swings for these devices. (c) Some devices exhibit SS down to 90$\,$mV/dec.}
\end{minipage}
\end{figure}

\begin{figure}[!h]
\vspace{0mm}
\begin{minipage} {\textwidth} %{\sminipagewidth}
\hspace{0cm}\includegraphics[width=16.5cm]{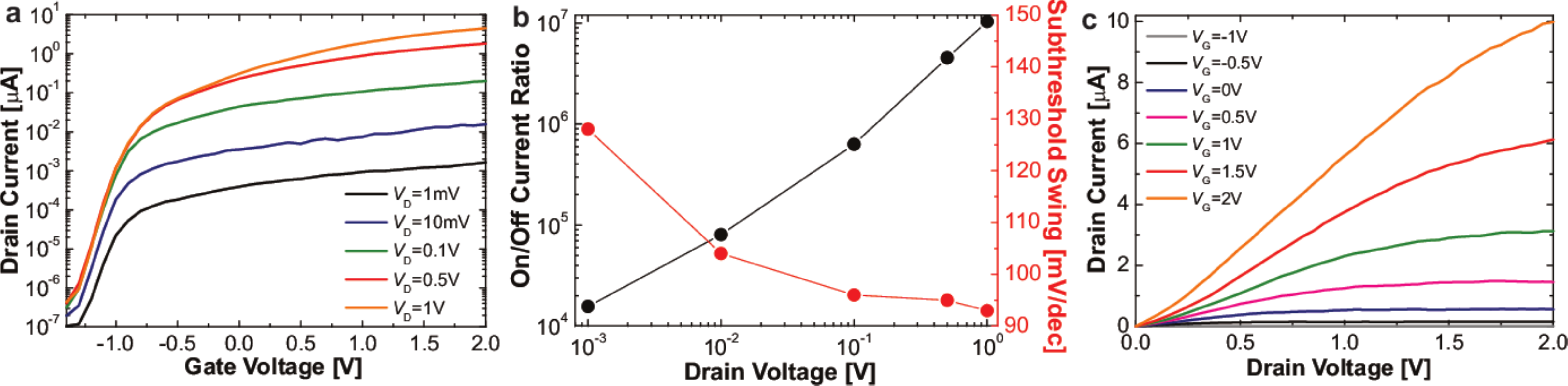} %???
\caption{\label{Fig.3} (a) $I_{\mathrm{D}}-V_{\mathrm{G}}$ characteristics measured for the best high current device. (b) On/off current ratio and SS extracted at different $V_{\mathrm{D}}$. The best performance of this device 
is achieved at $V_{\mathrm{D}}\,=\,$1$\,$V. (c) $I_{\mathrm{D}}-V_{\mathrm{D}}$ characteristics recorded on the same device exhibit saturation.}
\end{minipage}
\end{figure}

%Recently, even demonstration of one working transistor with tunnel-thin CaF$_2$ would be considered as an advance in the device technology. However, our advanced processing technique allowed us to
Using the process flow described above, we fabricated over 100 devices on two different CaF$_2$ substrates. In \Fig{Fig.2}a we show the gate transfer ($I_{\mathrm{D}}-V_{\mathrm{G}}$) characteristics measured for 50 devices from both substrates. The typical on-currents vary from 1$\,$nA to nearly 10$\mu$A, which is likely because of the non-homogeneous nature of the CVD MoS$_2$ film and different effective channel widths. At the same time, the measured on/off current ratios of some devices approach 10$^7$ (\Fig{Fig.2}b), which is excellent for back-gated MoS$_2$ FETs with a tunnel-thin gate insulator. Note that for the 
devices with overall lower currents, the measured on/off current ratios are probably underestimated due to the limited measurement resolution, which affects the off current. At the same time, the subthreshold swing (SS) values of most devices are smaller than 150$\,$mV/dec, while being close to 90$\,$mV/dec for some devices (\Fig{Fig.2}b,c). These values are among the best ever reported for back-gated MoS$_2$ FETs. Although in these prototypes 
SS$\,\sim\,$90$\,$mV/dec is achieved mostly for the devices with lower current (\Fig{Fig.2}c), several high-current devices also exhibit small SS values (e.g. \Fig{Fig.3}). As such, we believe that further optimization of CVD MoS$_2$ FETs on epitaxial fluorides and transition to more versatile configurations, such as top-gated devices and perhaps negative capacitance FETs with ferroelectric fluorides~\cite{RAVEZ97}, will lead to further improvements of these emerging devices. 

In \Fig{Fig.3}a we show typical $I_{\mathrm{D}}-V_{\mathrm{G}}$ characteristics measured for a device which simultaneously exhibits high drain currents and steep SS. The best transistor performance is achieved at $V_{\mathrm{D}}\,=\,$1$\,$V, with maximum measured on current about 5$\,\mu$A (or about 6$\,\mu$A/$\,\mu$m if normalized to the channel width), on/off current ratio close to 10$^7$ and SS as small as 93$\,$mV/dec (\Fig{Fig.3}b). %For higher $V_{\mathrm{D}}$ the off current becomes larger because of gate leakage through CaF$_2$, which degrades the device performance. 
The output ($I_{\mathrm{D}}-V_{\mathrm{D}}$) characteristics measured for different $V_{\mathrm{D}}$ (\Fig{Fig.3}c) also show promising behaviour with a large degree of current control and saturation. All these results confirm the promising nature of our devices and thus the high potential for further development.
\begin{figure}[!h]
\vspace{0mm}
\begin{minipage} {\textwidth} %{\sminipagewidth}
\hspace{1.cm}\includegraphics[width=15.0cm]{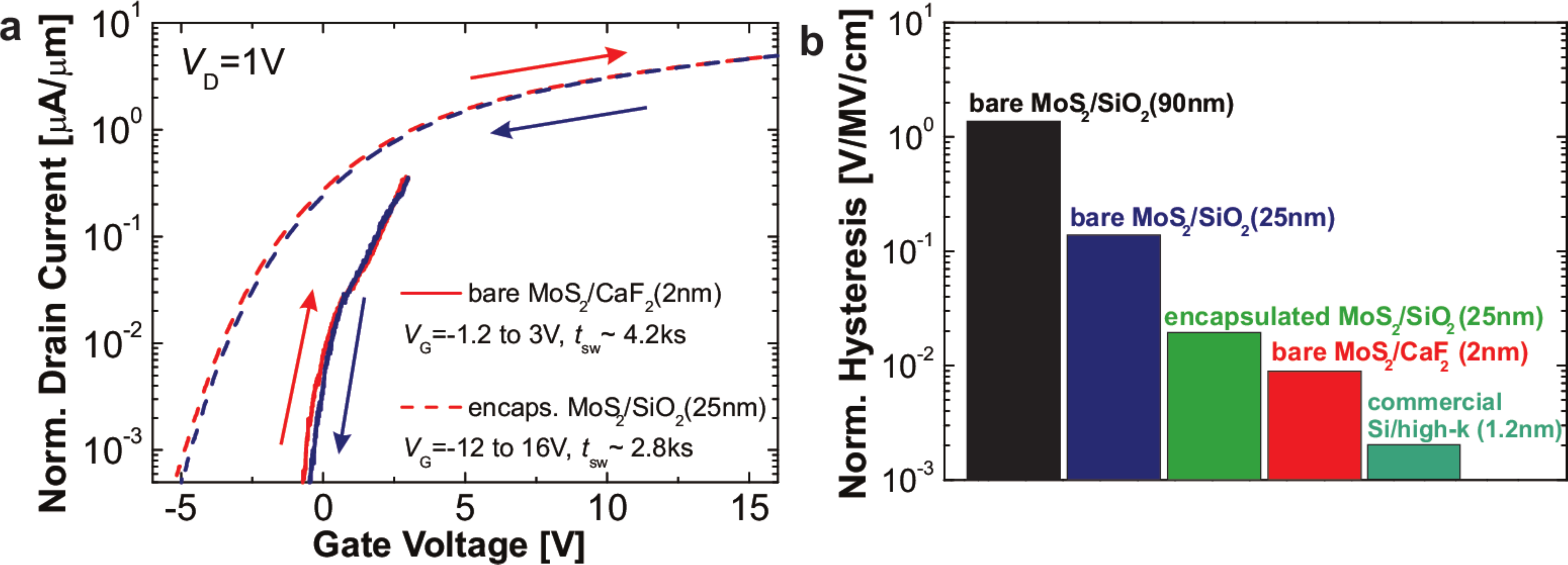} %???
\caption{\label{Fig.4} (a) Ultra-slow sweep $I_{\mathrm{D}}-V_{\mathrm{G}}$ characteristics measured for our MoS$_2$/CaF$_2$(2$\,$nm) FETs and Al$_2$O$_3$ encapsulated MoS$_2$/SiO$_2$(25$\,$nm) devices reported in our previous work~\cite{ILLARIONOV17C}. (b) For comparable sweep times, the hysteresis width has to be normalized to the insulator field factor $\Delta V_{\mathrm{G}}$/$d_{\mathrm{ins}}$. For our MoS$_2$/CaF$_2$(2$\,$nm) devices it is smaller compared to other 2D FETs and already close to commercial Si/high-k FETs.}
\end{minipage}
\end{figure}

Finally, we verify the electrical stability of our MoS$_2$/CaF$_2$ FETs by performing ultra-slow sweep hysteresis measurements with total sweep times $t_{\mathrm{sw}}$ of several kiloseconds. In \Fig{Fig.4}a we 
compare the $I_{\mathrm{D}}-V_{\mathrm{G}}$ characteristics measured for our MoS$_2$/CaF$_2$ devices with those of MoS$_2$/SiO$_2$ FETs with Al$_2$O$_3$ encapsulation reported in our previous 
work~\cite{ILLARIONOV17C}, which were found to exhibit the smallest hysteresis ever reported in 2D technologies. In addition to comparable values of $I_{\mathrm{D}}$ normalized to the channel width 
and considerably smaller SS for our MoS$_2$/CaF$_2$ FETs, we found that the hysteresis width near $V_{\mathrm{th}}$ is about 0.2$\,$V in both cases. However, for a fair comparison these values need to be 
normalized to the insulator field factor $\Delta V_{\mathrm{G}}$/$d_{\mathrm{ins}}$~\cite{ILLARIONOV17C}, with $\Delta V_{\mathrm{G}}$ being the width of $V_{\mathrm{G}}$ sweep range. While our devices with 
ultra-thin fluorite operate at considerably higher insulator fields of up to 15$\,$MV/cm (against 6.4$\,$MV/cm for the devices with thick SiO$_2$), their hysteresis stability is even better than for 
encapsulated MoS$_2$/SiO$_2$ FETs and close to commercial Si/high-k FETs, not to mention less advanced devices (\Fig{Fig.4}b). This confirms both the high electrical stability of CaF$_2$ as a gate insulator 
and the reduced number of slow insulator traps near the CaF$_2$/MoS$_2$ interface. The latter indicates that in addition to an excellent transistor performance, 2D FETs with CaF$_2$ are also more than competitive in terms of their reliability, which is a fundamental requirement for commercial applications. 

In summary, we have reported single-layer CVD MoS$_2$ FETs on an epitaxially grown CaF$_2$ insulator of record small 2$\,$nm thinness and demonstrated that our fully scalable technology allows fabrication of 
numerous transistors on a single chip. We have found that already these prototype devices can exhibit SS down to 90$\,$mV/dec and on/off current ratios up to 10$^7$, which are among the best values ever 
reported for back-gated devices. Furthermore, we have demonstrated that our devices are of very high electrical stability even for insulator fields of 10$\,$--$\,$15$\,$MV/cm and exhibit record small hysteresis ever 
measured for 2D devices. All this is due to the outstanding dielectric properties of CaF$_2$ and its good compatibility with MoS$_2$, which leads to a virtually defect-free quasi van der Waals interface between these 
materials. Together with the recent demonstration of epitaxial growth of 2D semiconductors on CaF$_2$(111)~\cite{VISHWANATH15,VISHWANATH18}, our findings present a breakthrough towards enabling ultra-scaled 
dielectric layers for next-generation 2D nanoelectronics. 

\section{Methods}

\textit{MBE growth of CaF$_2$ insulators} 

Ultra-thin CaF$_2$ layers were epitaxially grown on weakly doped single-crystal n-Si(111) substrates with $N_{\mathrm{D}}\,=\,10^{15}\,$cm$^{-3}$ and a misorientation of 5 to 10 angular minutes. Before the 
growth process, a protective oxide layer was formed after chemical treatment by following the procedure suggested by Shiraki~\cite{ISHIZAKA86}. Then, this layer was removed by annealing for 2 minutes at 
1200$^{\mathrm{o}}$C under ultra-high vacuum conditions ($\sim$10$^{-8}$$-$10$^{-7}$$\,$Pa). This allowed us to obtain an atomically clean 7$\times$7 Si(111) surface. After this, the CaF$_2$ film was grown 
on this surface by MBE at 250$^{\mathrm{o}}$C, which is known to be the optimum temperature to produce pinhole-free homogeneous fluorite layers~\cite{ILLARIONOV14D}. The deposition 
rate of fluorite measured by a quartz oscillator was about 1.3$\,$nm/min. The growth processes and crystalline quality of the CaF$_2$ layers were monitored using RHEED with an electron energy of 15$\,$keV (see 
the diffraction images shown in Supplementary Section 2).

\textit{Device fabrication} 

A single-layer MoS$_2$ film serving as a channel was grown on c-plane sapphire using the CVD process described by Dumcenco \textit{et al}~\cite{DUMCENCO15}. Namely, CVD growth was performed at atmospheric pressure and 
750$^{\mathrm{o}}$C using sulfur and MoO$_3$ as powder precursors and ultra-high-purity Ar as the carrier gas.

All lithography steps were done using E-Beam lithography. First we deposited SiO$_2$(5--10$\,$nm)/Ti/Au contact pads using sputtering. An isolation with the SiO$_2$ layer is required to minimize parasitic leakage currents through the 15$\,$--$\,$20$\,\mathrm{\mu}$m sized square electrodes, which have to be so large for a reliable contact with the probe. Then 7$\times$7$\,$mm CVD-grown MoS$_2$ films were transferred onto the CaF$_2$(111) substrate with pre-shaped isolated contact pads using the process suggested by Gurarslan \textit{et al}~\cite{GURARSLAN14}. In particular, we used a polystyrene film as a carrier polymer and dissolved it in toluene after the transfer process. The transferred MoS$_2$ film was subsequentily etched by reactive ion etching, in order to define the transistor channels with $L$ and $W$ between 400$\,$ and 800$\,$nm. Finally, the channels were 
contacted by e-beam evaporated Ti/Au pads deposited on top of MoS$_2$ in the contact areas. This second layer of Ti/Au was slightly extended to contact MoS$_2$ on top of the bare CaF$_2$ surface. 

\textit{TEM measurements} 

In order to achieve a high contrast in TEM measurements, the devices were covered by a 10 nanometers thick carbon layer deposited using sputtering. After this a TEM lamella preparation process was performed with a dual beam system. First a thicker granular platinum protective layer was deposited using a focused electron beam followed by a focused ion beam deposition. Then a TEM lamella was cut out along the channel of the device. Finally, the samples were examined using a TEM setup at the pressure of about 10$^{-5}$$\,$Pa. During the measurements, we recorded EELS spectra to verify the layer structure of our device.

\textit{Electrical characterization} 

Electrical characterization of our MoS$_2$/CaF$_2$ FETs consisted in the measurements of $I_{\mathrm{D}}-V_{\mathrm{G}}$ and $I_{\mathrm{D}}-V_{\mathrm{D}}$ characteristics. These measurements were conducted  
using a Keithley 2636 parameter analyzer in the chamber of a Lakeshore vacuum probestation ($\sim$5$\times$10$^{-6}$$\,$torr) at room temperature and in complete darkness. In order to correctly resolve the on/off current 
ratio, we used the autorange measurement mode. The hysteresis of the $I_{\mathrm{D}}-V_{\mathrm{G}}$ characteristics was investigated by doing double sweeps with a constant sweep rate. 

\begin{acknowledgement}
The authors thank for the financial support through the Austrian Science Fund FWF grant n$^\circ$ I2606-N30. T.M., D.K.P. and S.W. acknowledge financial support by the Austrian Science Fund FWF (START Y 539-N16) and the European Union (grant agreement No. 785219 Graphene Flagship). This work was partly supported by the Russian Foundation for Basic Research (grant No. 18-57-80006 BRICS$\_$t). We also gratefully acknowledge useful discussions with Mr. Markus Jech. Y.Y.I. is a member of Mediterranean Institute of Fundamental Physics (MIFP). 
\end{acknowledgement}

\section{Author contributions} 

Y.Y.I. introduced the idea of MoS$_2$ FETs with ultra-thin CaF$_2$ insulator, performed their characterization and prepared the manuscript. A.G.B. performed MBE growth of CaF$_2$ and provided the substrates. D.K.P. and S.W. fabricated MoS$_2$ FETs. M.S.-P. did TEM measurements. T.K. and M.T. contributed to preparation of figures. M.S-P. and A.S.-T. performed TEM measurements and sample preparation, respectively. M.I.V. performed quantitative analysis of gate leakage currents using tunnel models. M.W. programmed electrical measurements. N.S.S., T.M. and T.G. supervised this work. All authors regularly discussed the results and commented on the manuscript. 

\bibliography{./manuscript_arXiv_Illarionov.bbl}

\end{document}

\section{1. Tunnel leakage currents through CaF$_2$ and other insulators}

\begin{figure}[!h]
\vspace{-3mm}
\begin{minipage} {\textwidth} %{\sminipagewidth}
  \hspace{2cm}\includegraphics[width=12cm]{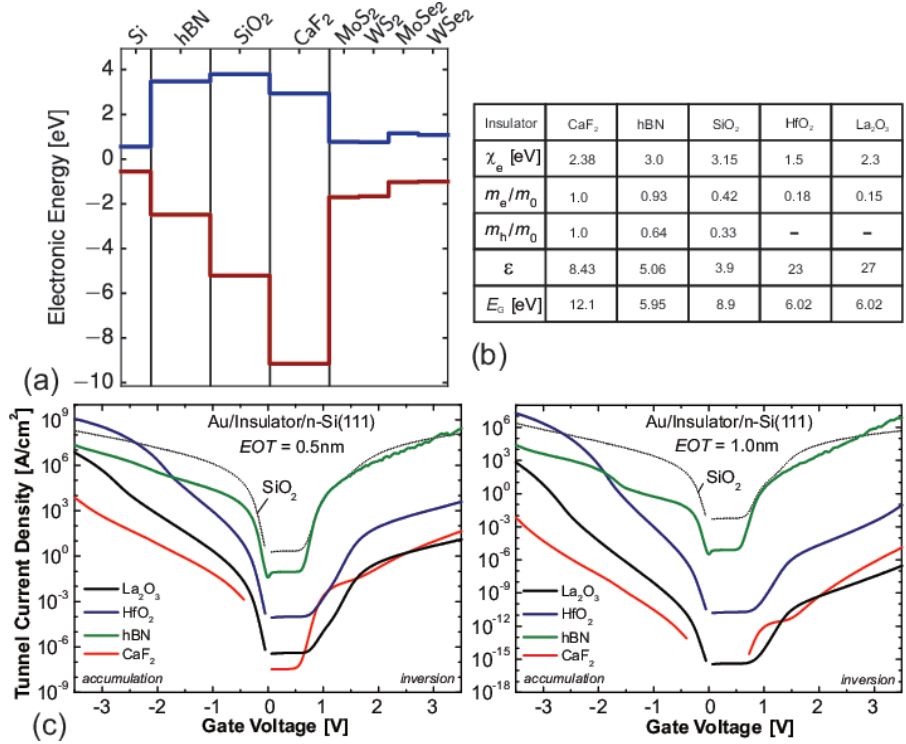} %???
\caption{(a) Band alignment of CaF$_2$, hBN and SiO$_2$ relative to several single-layer TMD semiconductors. (b) Dielectric parameters for different insulators. (c) Simulated tunnel leakages for a Au/insulator/n-Si(111) system with EOT$\,=\,$0.5$\,$nm (left) and EOT$\,=\,$1.0$\,$nm (right). The doping level of the substrate was $N_{\mathrm{D}}$$\,=\,$10$^{15}$$\,$cm$^{-3}$. In order to maintain a proper comparison of different insulator systems, the condition $\chi_{\mathrm{m}}\,=\,\chi_{\mathrm{e}}\,+\,0.25\,$eV is used for the band offset with Au.   } 
\end{minipage}
\end{figure}

In Fig.$\,$S1a we show the energetic alignment of CaF$_2$, hBN and SiO$_2$ relative to Si and several single-layer TMD semiconductors obtained using the band offsets known from literature~\cite{ILLARIONOV10,RASMUSSEN15}. Clearly, CaF$_2$ has a considerable advantage in terms of bandgap, which is twice as large as that of hBN. This fact alone should lead to considerably smaller tunnel currents for the same equivalent oxide thickness (EOT). The other parameters which affect tunnel leakages are the dielectric constant $\varepsilon$, the electron and hole effective masses ($m_{\mathrm{e}}$ and $m_{\mathrm{h}}$, respectively) and the conduction band offset with the Si substrate $\chi_{\mathrm{e}}$. The values known from the literature~\cite{GEICK66,HAYES74,ROBERTSON00,HOU03,MIRANDA05,VEXLER05,RASMUSSEN15} for CaF$_2$, hBN, SiO$_2$ and high-k oxides HfO$_2$ and La$_2$O$_3$ are summarized in Fig.$\,$S1b. Using these parameters and our well-established modeling technique~\cite{VEXLER09,VEXLER10A}, we modeled the tunnel currents through the Au/insulator/n-Si(111) system. Our approach is based on a WKB calculation of the tunneling probability and accounts for transverse momentum conservation~\cite{VEXLER09,VEXLER10A,ILLARIONOV11C} in the case of crystalline CaF$_2$ on Si(111). As for the case of oxide insulators and hBN, the tunneling current is supposed to occur as if silicon were a direct-band semiconductor (otherwise a huge difference between the Si(111) and Si(100) cases would have been predicted, which is never observed). The results obtained for EOT$\,=\,$0.5$\,$nm and EOT$\,=\,$1.0$\,$nm are shown in Fig.$\,$S1c. Clearly, in both cases the tunnel leakage through CaF$_2$ is even smaller than for high-k oxides, not to mention hBN, which is closer to SiO$_2$ rather than to high-k materials~\footnote{Note that for hBN we use the best case values of $\chi_{\mathrm{e}}\sim$3$\,$eV~\cite{RASMUSSEN15} and $\varepsilon\,=\,$5.06~\cite{GEICK66}. However, this material is not perfectly parameterized. For instance, other suggested $\chi_{\mathrm{e}}$ values are about 1$\,$eV smaller~\cite{LEE11A} than in Fig.$\,$S1a and an alternative $\varepsilon$ would be 3.76~\cite{LATURIA18}. As such, our model provides the results for minimum possible tunnel leakages through hBN.}. Together with a defect-free substrate and a quasi van der Waals interface with 2D semiconductors, small leakages through CaF$_2$ layers with sub-1$\,$nm EOT make this material the most promising for scaled next-generation 2D devices. 

\section{2. RHEED control of CaF$_2$ epitaxy}

\begin{figure}[!h]
\vspace{-3mm}
\begin{minipage} {\textwidth} %{\sminipagewidth}
  \hspace{1cm}\includegraphics[width=15.0cm]{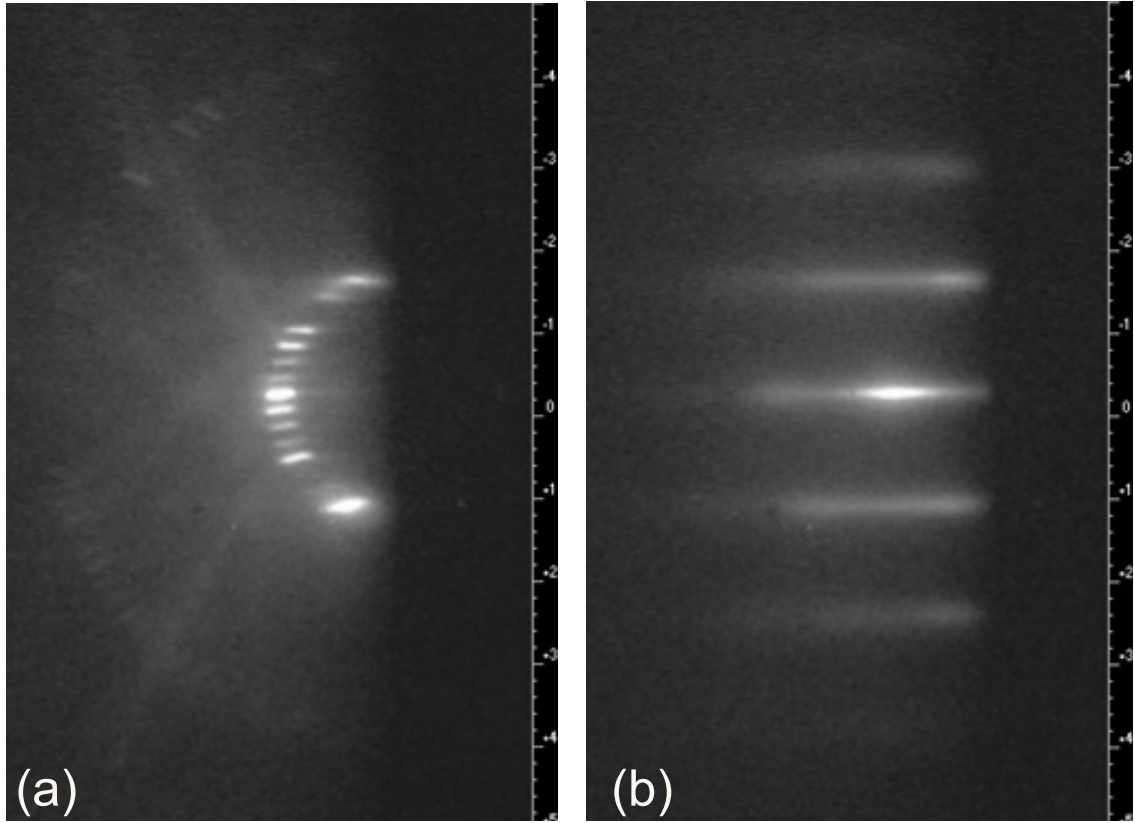} %???
\caption{RHEED patterns of the Si(111) surface showing the 7$\times$7 superstructure (a) and a 2$\,$nm thick CaF$_2$ layer grown at 250$^{\mathrm{o}}$C (b). Azimuth of the electron beam is [1$\bar{1}$0]. } 
\end{minipage}
\end{figure}

The MBE growth process and the crystalline quality of the CaF$_2$ layers were monitored using reflection high energy electron diffraction (RHEED)~\cite{SOKOLOV92} with an electron energy of 15 keV. In Fig.$\,$S2 we show the typical RHEED patterns obtained for atomically clean Si(111) before the beginning of fluorite epitaxy (a) and from a deposited CaF$_2$ layer at the final stage of the MBE process (b). The presence of distinct reflections in the 
CaF$_2$ pattern indicates its high crystalline quality even at a comparably low MBE growth temperature of 250$^{\mathrm{o}}$C. Note that although the use of a higher growth temperature (e.g. 750$^{\mathrm{o}}$C) would further improve the crystalline quality of fluorite, the layer would then become non-homogeneous with a number of pinholes. 

\section{3. Geometry of our MoS$_2$/CaF$_2$ FETs}

\begin{figure}[!h]
\vspace{-3mm}
\begin{minipage} {\textwidth} %{\sminipagewidth}
  \hspace{2cm}\includegraphics[width=13.5cm]{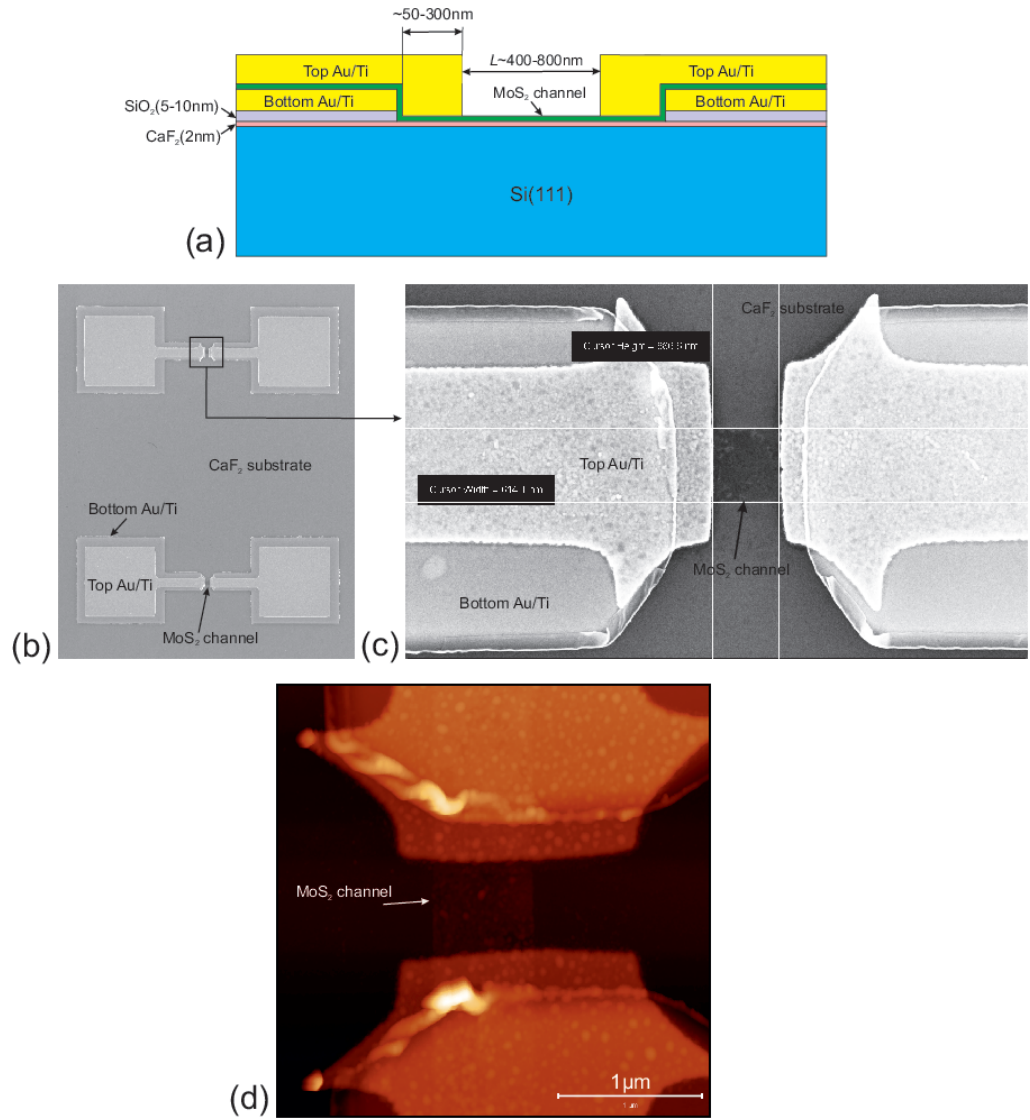} %???
\caption{(a) Schematic cross-section of our MoS$_2$/CaF$_2$ FETs. (b) SEM image of two devices with contact pads. (c) SEM image of the channel area with marked channel dimensions. (d) AFM image of the channel area. } 
\end{minipage}
\end{figure}

In Fig.$\,$S3a we show a schematic cross-section of our MoS$_2$/CaF$_2$ FETs. Reliable contact of the test devices with the probes requires contact pads as large as 15$\,$--$\,$20$\,$$\mu$m. However, the CaF$_2$ layer may contain 
some thickness fluctuations and local pinholes with the depth of several angstroms, which may locally increase the tunnel leakage by orders of magnitude. The density of these pinholes typically depends on the quality of the Si(111) substrate (e.g. misorientation angle, roughness, etc.) and the MBE growth parameters. It is commonly assumed that for a well adjusted MBE process of CaF$_2$ on Si(111) there should be less than one pinhole per 
100$\,$$\mu$m$^2$. As such, the probability of having some pinholes under the 225$\,$--$\,$400$\,$$\mu$m$^2$ sized pads is quite high, which may lead to a smaller number of functional devices. Thus, in order to reduce parasitic 
leakages, the contact pads received some additional insulating layer with 5$\,$--$\,$10$\,$nm thick SiO$_2$. At the same time, the contact of the SiO$_2$ layer with the MoS$_2$ film is completely avoided, which is necessary to block possible charge trapping events at the MoS$_2$/SiO$_2$ interface and thus make the transistor characteristics more stable. As such, within the contact area, the MoS$_2$ film is sandwiched between two Ti/Au layers, in which a thin (few nanometers) Ti layer is used as an adhesion layer. A typical view of our devices obtained using a scanning electron microscope (SEM) is shown in Fig.$\,$S3b. A detailed SEM image of the channel area (Fig.$\,$S3c) allows to estimate the channel dimensions. For our devices both $L$ and $W$ are typically between 400 and 800$\,$nm. However, the CVD MoS$_2$ film contains some imperfections, which may lead to different effective channel widths for different devices. Within the channel area, the MoS$_2$ film is just on top of the CaF$_2$ insulator. As shown in Fig.$\,$S3d, the channel area of our devices can be also nicely resolved using atomic-force microscope (AFM). 

\section{4. Estimation of the effective gate insulator thickness in our MoS$_2$/CaF$_2$ FETs}

\begin{figure}[!h]
\vspace{-3mm}
\begin{minipage} {\textwidth} %{\sminipagewidth}
  \hspace{1cm}\includegraphics[width=15cm]{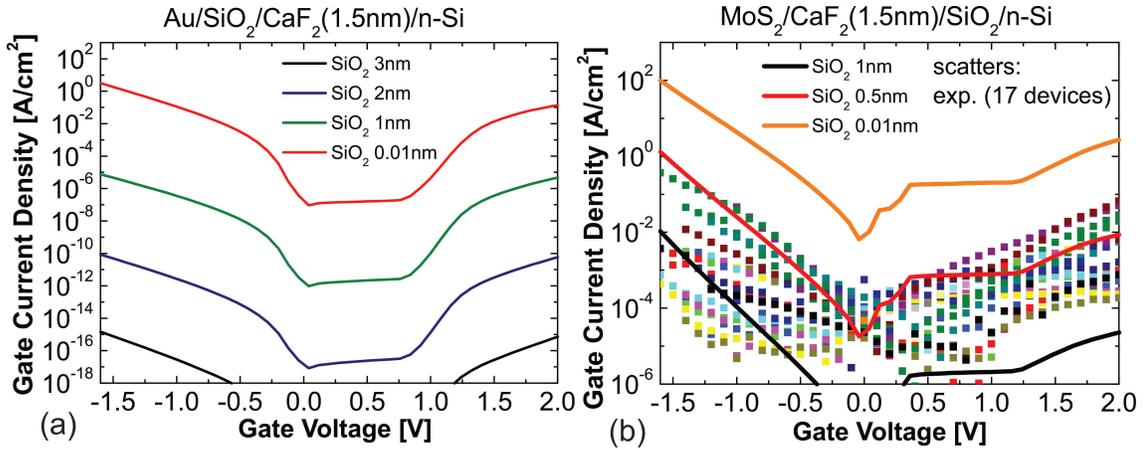} %???
\caption{(a) Simulated tunnel leakages for Au/SiO$_2$/CaF$_2$/n-Si(111) structures representing the contact pads of our MoS$_2$ FETs. (b) Simulated tunnel leakages for MoS$_2$/CaF$_2$/SiO$_2$/n-Si(111) structures representing the channel and experimental data measured for 17 devices. It appears that the effective thickness of our gate insulator is about 2$\,$nm, which roughly corresponds to 1.5$\,$nm of CaF$_2$ and 0.3$\,$--$\,$0.7$\,$nm of SiO$_2$ due to oxygen penetration.} 
\end{minipage}
\end{figure}

Taking into account that thickness fluctuations may be present in tunnel-thin CaF$_2$ films~\cite{ILLARIONOV14D}, the effective thickness of the insulator can be different from the physical thickness which can be visually seen in TEM images. For a qualitative estimation of the effective thickness, we theoretically model the tunnel leakages and compare the results with our experimental data.
First we model the tunnel leakages through Au/SiO$_2$/CaF$_2$/n-Si structures which form our contact pads. In Fig.$\,$S4a we show that already for a SiO$_2$ thickness of 3$\,$nm the leakage becomes negligible, while an increase of the insulator thickness by 1$\,$nm leads to a decrease of the tunnel current by  about 5 orders of magnitude. As such, our 15--20$\,\mu$m sized contact pads with 5$\,$--$\,$10$\,$nm SiO$_2$ can not contribute to the measured tunnel leakage. This means that the measured gate current is mostly given by the tunnel leakage within the channel area. The corresponding modeling results are shown in Fig.$\,$S4b. Based on our TEM measurements, we assume that the physical thickness of CaF$_2$ is 7$\,$ML (1$\,$ML=0.315$\,$nm), which is about 22$\,\mathrm{\AA{}}$. The root mean square (rms) amplitude of fluctuations $\sigma$ measured for this fluorite thickness using atomic-force microscope (AFM)~\cite{ILLARIONOV14D} is about 2.7$\,\mathrm{\AA{}}$. This gives an effective thickness ($d_{\mathrm{eff}}\,=\,d_{\mathrm{phys}}\,-\sigma^2$, all values in angstroms) of CaF$_2$  $d_{\mathrm{eff}}\,\sim\,$1.5$\,$nm. In agreement with our previous observations~\cite{ILLARIONOV15DS}, ambient storage of CaF$_2$ films grown on Si leads to some oxidation at the CaF$_2$/Si interface, since oxygen is able to penetrate through the thinnest places in the CaF$_2$ layers. As a result, a thin SiO$_2$ layer is formed underneath CaF$_2$. While previously we observed this oxidation as a decrease of the tunnel currents after several months of device storage~\footnote{Our MoS$_2$ FETs were fabricated after 8--12 months following the epitaxial growth of CaF$_2$.}, in this study the presence of the thin SiO$_2$ layer was confirmed by TEM measurements (see Fig.$\,$S5c). As such, in our model we assume MoS$_2$/CaF$_2$/SiO$_2$/n-Si structures and vary the thickness of SiO$_2$. Fitting of our modeling results with the tunnel gate currents measured for numerous devices in Fig.$\,$S4b allows us to conclude that the effective thickness of the SiO$_2$ oxidation layer is about 0.3$\,$--$\,$0.7$\,$nm. Therefore, the total effective thickness of the gate insulator in our MoS$_2$ FETs is about 2$\,$nm. Note that the oxidation layer underneath CaF$_2$ does not affect the device performance, since it is far away from defect-free CaF$_2$/MoS$_2$ interface. %Even more, this layer leads to some improvement of the insulator and thus leads to an improvement of the on/off current ratio. 

\section{5. EELS analysis and sample degradation during TEM }

\begin{figure}[!h]
\vspace{-3mm}
\begin{minipage} {\textwidth} %{\sminipagewidth}
   \hspace{1.5cm}\includegraphics[width=14cm]{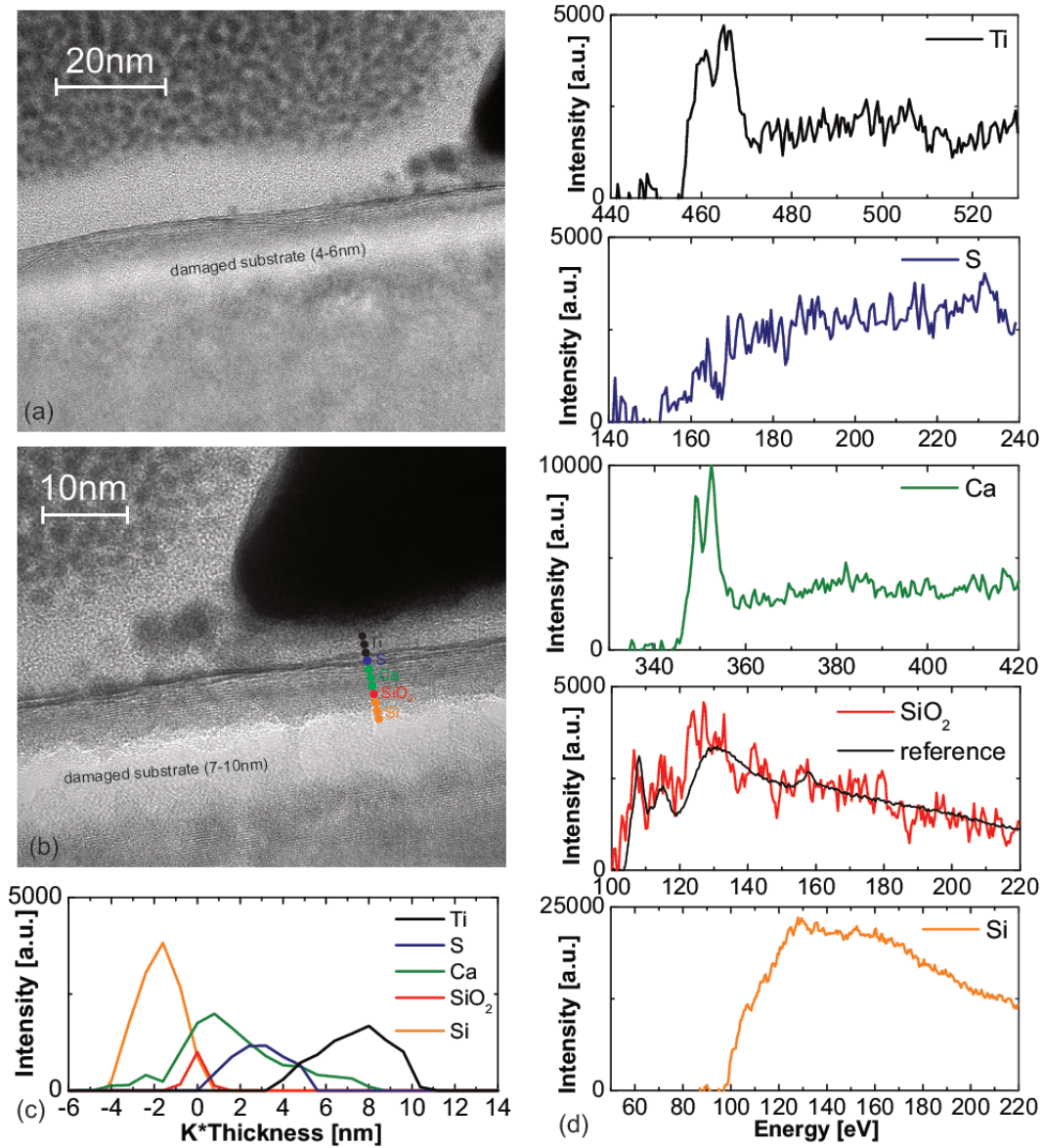} %???
\caption{(a) TEM image obtained at the beginning of our measurements. The Si substrate is damaged by TEM a few nanometers below the CaF$_2$/Si interface. (b) TEM image obtained after 2 more minutes. The damage of the Si substrate is progressing. However, the features of the Si/CaF$_2$/MoS$_2$/Ti structure can be clearly resolved. (c) The cross-sectional EELS spectrum measured along the line marked in (b) confirms the layer alignment and suggests the presence of a thin SiO$_2$ layer at the Si/CaF$_2$ interface. (d) EELS spectra of each particular material detected.} 
 \end{minipage}
 \end{figure}

In Fig.$\,$5Sa we show the TEM image measured at the beginning of our study. We observe that already during the first measurement the TEM irradiation leads to a partial damage of Si substrate a few nanometers below the Si/CaF$_2$ interface. The origin of this issue is not clear yet, but could be a chemical modification of Si (e.g. by acetone penetrating through CaF$_2$ during MoS$_2$ transfer). Additionally, this damaged region is less stable with respect to irradiation. In particular, this damage appears only in the areas where MoS$_2$ is right on top of CaF$_2$. However, it does not appear away from the channel, where CaF$_2$ is isolated by sputtered SiO$_2$. After two more minutes of irradiation and focusing of the beam, the damaged area expands considerably (Fig.$\,$5Sb). Nevertheless, the Si/CaF$_2$ and CaF$_2$/MoS$_2$ interfaces remained stable enough to allow for recording an electron energy loss spectrometry (EELS) line scan employing the scanning mode of the TEM. In this situation a low dose electron beam is focused and scanned along the line shown in Fig.$\,$5Sb. In each position an EELS spectrum is recorded, thus having a spatial resolution of 0.5$\,$nm. As shown in Fig.$\,$5Sc, the EELS profile clearly confirms the layered structure of our device, though the measured thickness is broadened by an unknown factor $K$ due to the non-planar surface. Also, the presence of the thin SiO$_2$ layer at the Si/CaF$_2$ interface is clearly confirmed by observation of the fine structure of the S-L ionization edge at the respective position of the linescan. This thin oxide layer appears as a result of several months of storage of the Si/CaF$_2$ substrates before device fabrication and has been accounted for in the model above. Interestingly, in the damaged area the Si signal drops, which confirms the removal of the material by TEM irradiation. The EELS spectra of all the materials detected are shown in Fig.$\,$5Sd and agree well with their reference shapes from the EELS atlas.  

\begin{figure}[!h]
\vspace{-3mm}
\begin{minipage} {\textwidth} %{\sminipagewidth}
  \hspace{0cm}\includegraphics[width=16.5cm]{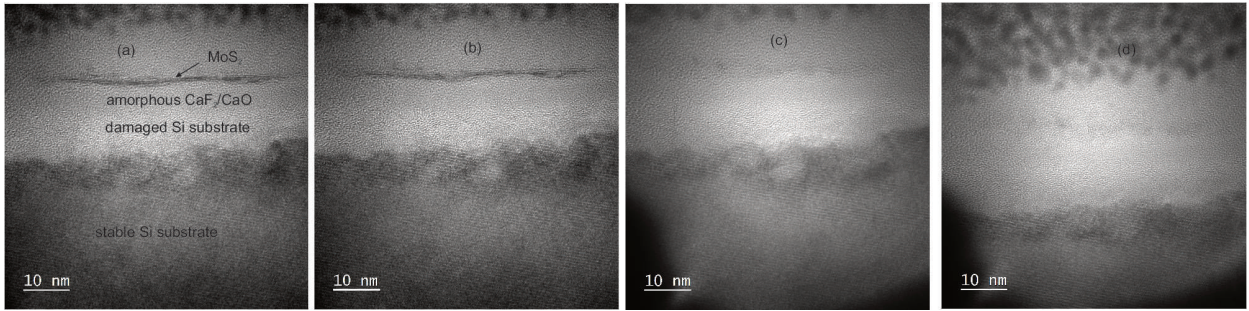} %???
\caption{TEM images of the channel measured with high electron dose rate. Images a-d are taken within 4 minutes, which is enough to destroy the channel completely.} 
\end{minipage}
\end{figure}

Finally, we record TEM images of the channel with an illuminated area of about 75$\,$nm in diameter, a beam current of 26$\,$nA giving an electron dose rate of 3.7$\times$10$^7$ electrons/s/nm$^2$, which is at least an order of magnitude larger than for all our previous TEM measurements. As shown in Fig.$\,$S6, already the first measurement with such a high electron dose rate leads to a complete amorphisation of CaF$_2$ and partial damage of MoS$_2$. A further increase of irradiation time leads to a severe damage of the Si substrate, and after about 4 minutes all layers are completely destroyed (Fig.$\,$S6d). These results confirm that TEM is in general destructive for MoS$_2$/CaF$_2$ FETs and reasonable results can be obtained only with moderate electron dose rates. 

\bibliography{./si_arXiv_Illarionov.bbl}